**Oxide two-dimensional electron gas with high mobility at room-temperature**

Kitae Eom,[1] Hanjong Paik,[2,3] Jinsol Seo,[4] Neil Campbell,[5] Evgeny Y. Tsymbal,[6] Sang Ho Oh,[4] Mark Rzchowski,[5] Darrell G. Schlom,[2, 7, 8] and Chang-Beom Eom*,[1]

[1]Department of Materials Science and Engineering, University of Wisconsin-Madison, Madison, WI 53706, USA

[2]Department of Material Science and Engineering, Cornell University, Ithaca, New York 14853, USA

[3]Platform for the Accelerated Realization, Analysis, and Discovery of Interface Materials (PARADIM), Cornell University, Ithaca, New York 14853, USA

[4]Department of Energy Science, Sungkyunkwan University (SKKU), Suwon, 16419, Republic of Korea

[5]Department of Physics, University of Wisconsin, Madison, WI 53706, USA

[6]Department of Physics and Astronomy, University of Nebraska, Lincoln, Nebraska 68588, USA

[7]Kavli Institute at Cornell for Nanoscale Science, Ithaca, New York 14850, USA

[8]Leibniz-Institut für Kristallzüchtung, Berlin 12489, Germany

Correspondence should be sent to eom@engr.wisc.edu.

**The prospect of 2-dimensional electron gases (2DEGs) possessing high mobility at room temperature in wide-bandgap perovskite stannates is enticing for oxide electronics, particularly to realize transparent and high-electron mobility transistors. Nonetheless only a small number of studies to date report 2DEGs in $BaSnO_3$-based heterostructures. Here, we report 2DEG formation at the $LaScO_3/BaSnO_3$ (LSO/BSO) interface with a room-temperature mobility of 60 $cm^2/V·s$ at a carrier concentration of $1.7 \times 10^{13}$ $cm^{-2}$.**

**This is an order of magnitude higher mobility at room temperature than achieved in $SrTiO_3$-based 2DEGs. We achieved this by combining a thick BSO buffer layer with an *ex-situ* high-temperature treatment, which not only reduces the dislocation density but also produces a $SnO_2$-terminated atomically flat surface, followed by the growth of an overlying BSO/LSO interface. Using weak-beam dark field imaging and in-line electron holography technique, we reveal a reduction of the threading dislocation density, and provide direct evidence for the spatial confinement of a 2DEG at the BSO/LSO interface. Our work opens a new pathway to explore the exciting physics of stannate-based 2DEGs at application-relevant temperatures for oxide nanoelectronics.**

Two dimensional electron gases (2DEGs) at oxide interfaces have attracted significant attention in both fundamental research as well as potential device applications. Among them, the heterointerface between LaAlO$_3$ and SrTiO$_3$ is the most studied prototype system. Fascinating physical phenomena including magnetism,[1] superconductivity,[2,3] strong spin-orbit interactions,[4,5] and exotic quantized transport.[6,7] have been reported. Unfortunately, despite extensive work on STO-based 2DEG heterostructures with overlayers such as LaAlO$_3$,[8] LaTiO$_3$,[9] NdGaO$_3$[10] and γ-Al$_2$O$_3$,[11] room-temperature mobilities of interfacial 2DEGs are poor (e.g., < 10 cm$^2$/V·s). This arises from the nature of electron states in the narrow Ti $d$-bands that host the 2DEG in STO, their interaction with the crystalline lattice, and multiple interband scattering channels due to the degenerate $t_{2g}$ orbital symmetry.[12] This has stymied wide-ranging room-temperature 2DEG applications in these systems.

One route toward higher room-temperature interfacial 2DEG mobilities is to move away from STO to a non-polar oxide with more dispersive $s$ or $p$ orbital-based conduction bands to provide highly mobile carriers at room temperature. BaSnO$_3$ (BSO) has gained significant attention in theory and experiment as an alternative interfacial 2DEG host material.[13-15] The conduction band of BSO is composed of highly dispersive non-degenerate $s$-orbitals with a large band width and a low effective mass. Additionally, the interband scattering channel can be turned off from the isotropic $s$ orbitals like conduction band structure, resulting in a longer lifetime for the charge carrier.[8] Therefore, the BSO based 2DEG's mobility at room temperature is predicted to be two orders of magnitude higher than that of STO based 2DEGs in a structurally perfect BSO host.[16-18]

The sensitivity of carrier mobility and free carrier concentration to structural and point defects makes achieving a high mobility 2DEG in BSO challenging. One such structural defect is the high density of dislocations in typical BSO thin films (more than $10^9$ cm$^{-2}$).[22-25] These threading dislocations originate from the large lattice mismatch between BSO and commercially available perovskite substrates (ranging from −5.2% (SrTiO$_3$) to −2.4% (PrScO$_3$)). Because of such defects, the mobility of electrons produced by La doped thin films of BaSnO$_3$ [22] has not yet exceeded that of La-doped BaSnO$_3$ bulk single crystals.[16,17] Another compromising factor is the formation of complex point defects during BSO film growth, which act as additional electron traps or scattering sites.[22] Finally, producing an interfacial 2DEG in a BSO-based bilayer heterostructure requires proper band structure alignment and a good epitaxial match between the top oxide layer and BSO.

For these and other reasons, only a few experimental studies have pursued BSO-based 2DEG formation, using either modulation doping or polarization doping.[19-21] A recent report has claimed an interfacial conductivity improvement at the LaInO$_3$/La-BaSnO$_3$ interface, akin to the polar catastrophe polarization doping scenario.[19,20] La-doped (0.3%) BSO channels were, however, used to compensate a high density of defect states at the LaInO$_3$/BaSnO$_3$ interface. Prakash et al. reported a modulation doping approach utilizing La-SrSnO$_3$/BaSnO$_3$ hetero-structures, where electrons from the La-doped SrSnO$_3$ side spill over into an undoped BSO layer.[21] They found that the modulated electrons spread over 3-4 unit cells toward the BSO layer, even though La-doped SrSnO$_3$ layer is the more dominant conducting path than the modulated BSO layer. Nonetheless, the reported 2DEG's carrier behaviors did not unambiguously demonstrate 2DEG confinement at the hetero-interface.

Here we report a BSO-based highly mobile interfacial 2DEG, where we have overcome the electronic alignment and defect density issues discussed above. We incorporate LaScO$_3$ (LSO) as a top polar layer epitaxially registered with a non-polar host BSO thin film. LSO is orthorhombic, but we give pseudocubic LSO thicknesses throughout this manuscript. Similar to the case of LAO/STO, the LSO/BSO interface has a band alignment that facilitates 2DEG formation[13] (see **Fig. 1**a, LSO/BSO band diagram) while possessing excellent structural coherency. The mismatch strain is less than 1.6 %, as is evident in the atomic model of the LSO/BSO interface shown in Fig. 1b. The LSO/BSO interface hosts a polar discontinuity (Figure 1a) [13], facilitating 2DEG formation at atomically sharp LaO/SnO$_2$ interfaces. We dramatically reduced the BSO defect densities by first growing a thick BSO layer by pulsed-laser deposition (PLD), annealed it at high temperature *ex-situ*, and then continued the growth of BSO by MBE to form the desired interface between BSO and the polar LSO layer. Electronic transport measurements revealed an insulator-to-metal transition at a threshold thickness ($t_c$) of about 4 unit cells (u.c.), consistent with the polar catastrophe model,[13] and a room-temperature mobility as high as 60 cm$^2$/Vs with a $1.7 \times 10^{13}$ cm$^{-2}$ carrier density. In-line electron holography showed negative charges confined to the LSO/BSO interface. TEM analysis verified a reduced dislocation density resulting from our synthesis approach. Complete structural and morphological analysis demonstrates high crystalline quality. This first demonstration of a high mobility BSO-based 2DEG provides a fascinating platform for exploring transparent conducting oxide electronic devices and the physics of two-dimensional *s*-orbital systems.

Our approach to minimizing dislocation density starts with a PLD-grown thick BSO buffer layer (about 550 nm) (Experimental Section), as shown in **Figure 2**. The thick BSO layer is leached in water for 15 sec and then *ex-situ* annealed in oxygen at 1150 ℃ for 2 hours (which we refer to as the BSO pseudo-substrate hereafter)[26] before MBE regrowth. We note that undoped BSO buffer layers about half this thickness were used to achieve the highest La-doped BaSnO$_3$ single film electron mobilities to date.[22] We summarize the structural analysis of the BSO pseudo-substrate before and after thermal treatment in Figure S1. This treatment not only reduces the dislocation density,[27] but also produces SnO$_2$-terminated atomically flat surfaces. The full width at half maximum (FWHM) of the BSO 002 peak's rocking curve is 0.013 degrees after treatment (Figure S1d). The in-plane and out-of-plane lattice constants of the BSO pseudo-substrate obtained from the reciprocal space maps of 103 BSO peaks (Figure

S1f) are 4.112 Å and 4.118 Å, respectively, indicating an almost fully relaxed state.[28] Atomic force microscope images show an atomically flat surface of the BSO pseudo-substrate with single unit cell steps of 0.4 nm (Figure 2c).

After loading the BSO pseudo-substrate into the MBE, we grew a 45 nm thick BSO layer in an adsorption-controlled regime, followed by a 10 u.c. thick LSO layer (Figure 2d) (Experimental Section; Figure S2). Using this method, we form the LSO/BSO interface away from the air-exposed BSO surface, yet capitalize upon the benefits of the pseudo-substrate with lowered threading dislocation density and desired $SnO_2$-terminated surface. Further, the 2DEG interface is produced in the MBE-grown portion of the structure known to produce high-mobility BSO layers. A reciprocal space map around the (103) STO substrate peak shows that the LSO film is fully coherent with respect to the underlying BSO film (Figure 2d).

**Figure 3** shows the temperature-dependent transport properties of the 10 u.c. thick MBE LSO/BSO 2DEGs grown on the BSO pseudo-substrate (red squares). It is compared to a control sample of 10 u.c. LSO/BSO (60 nm) 2DEGs grown without the BSO pseudo-substrate (directly grown on STO (001) substrate by MBE, blue circles). Both systems show semiconducting-like features across the entire measurement temperature range between 100 K and 400 K (Figure 3a). The carrier density of the LSO/BSO 2DEG grown on the BSO pseudo-substrate decreases monotonically over the temperature range and shows a 4.5 times higher carrier density than the control sample (Figure 3b). In addition, the LSO/BSO 2DEG grown on the BSO pseudo-substrate has 6.8 times lower sheet resistance and 3.4 times higher mobility (Figure 3c) than the control sample. Notably, the highest 2DEG mobility at room temperature is 60 $cm^2/V·s$ with carrier concentration $1.7 \times 10^{13}$ $cm^{-2}$. This is an order of magnitude higher 2DEG mobility at room temperature than in STO-based 2DEGs.[8-11,29]

Our TEM measurements quantify that the dislocation density of LSO/BSO

heterostructures grown on the BSO pseudo-substrate is reduced from the control sample, which underlies the observed enhanced room-temperature mobility and carrier density. Misfit dislocations are known to be prevalent in BSO grown on STO substrate due to the large lattice mismatch. We found that threading dislocations propagate along the film growth direction (**Figure 4**a and b) from the BSO/STO interface to the LSO layer (Figure S3). Figure 4a shows the TEM weak-beam dark-field images of the control sample (10 u.c. LSO/BSO 60 nm grown on a STO (001) substrate). We evaluate a threading dislocation density on the order of ~$10^{11}$ cm$^{-2}$ (Figure 4c) from the image analysis as shown in Figure S4 (Experimental Section; Supporting Information). The misfit dislocations at the BSO film-STO substrate interface are denoted by red arrows (Figure S5b). Notably, the LSO/BSO heterostructure grown on the BSO pseudo-substrate showed a lower dislocation density of $4.1 \times 10^{10}$ cm$^{-2}$ (Figure 4e), less than half of the density observed when grown directly on an STO (001) substrate (Figure 4c). This is because the PLD grown BSO film was annealed at high temperature which has been shown to cause annihilation of threading dislocations.[24,27] The dislocation density of the BSO layer grown on the BSO pseudo-substrate is very similar to that of the BSO pseudo-substrate (Figure 4d). This provides less charge trapping and scattering, consistent with the increased carrier density and mobility at room temperature for the LSO/BSO heterostructure grown on the BSO pseudo-substrate.

We used in-line electron holography to quantify 2DEG confinement near the interface, and to support the LSO thickness-dependent electrical transport properties of the LSO/BSO interface (Experimental Section; Supporting Information). For these measurements we grew the LSO layer by PLD on an MBE-grown 90 nm thick BSO layer grown on a STO (001) substrate (Supporting information). All BSO surface were SnO$_2$ terminated with single unit cell steps, achieved by a water leaching treatment.[26] The LSO film thickness was controlled by reflection high-energy electron diffraction (RHEED) intensity oscillations (Figure S6).

Atomically resolved STEM-energy-dispersive X-ray spectroscopy (EDS) elemental mapping across the interface verified the LaO/SnO$_2$ termination of the LSO/BSO interface (**Figure 5**a) (Experimental section and Supporting information). Different LSO thickness heterostructures established a critical thickness ($t_c$) of 4 unit cells for conductivity (Figure S7), consistent with a polar catastrophe interpretation[30]. In-line electron holography results of a 4 u.c. thick LSO/BSO interface (Figure S8) shows no significant net charge density near the interface, while those of the 10 u.c. LSO/BSO (Figure 5b) show a 2D charge density equivalent to $5 \times 10^{21}$ cm$^{-3}$, distributed with a peak 1.5 nm below the LSO/BSO interface before quickly decaying to zero around 5 nm below the interface. There is no long tail of electron density extending deep into the BSO side.

This depth dependence is in contrast that obtained from in-line holography results of LAO/STO heterostructures, which show a charge confinement within 1.5 nm below the interface and a maximum electron density of $5 \times 10^{21}$ cm$^{-3}$ located 0.5 nm below the interface.[31] We attribute these differences primarily to the lower dielectric constant of BSO compared to STO.[32,33] A secondary reason could be the difference between Sn 5$s$ orbitals in BSO and Ti 3$d$ orbitals in STO from which the 2DEG electron states arise. A related broadening of the 2DEG extent (4.5 nm) was reported in the (111) oriented LAO/STO 2DEGs.[31] (111)-oriented LSO/STO 2DEGs due to orbital orientation in STO. It is worth noting that the carrier density of LSO/BSO 2DEGs obtained by a Hall effect measurement is a much smaller value than that measured by inline holography. This discrepancy may come from the fact that the Hall effect measurement is sensitive to mobile charge carriers, whereas inline holography reflects the total charge density, including both mobile and localized charges.[31]

Our large room-temperature mobility of 60 cm$^2$/V·s (Figure 3c), is still lower than that predicted theoretically.[13] Interfacial cation intermixing at the LSO/BSO interface could reduce the carrier mobility, a common phenomenon in oxide interfaces. The STEM-EDS mapping results shown in Figure 5a, however, do not indicate significant cation intermixing at the LSO/BSO interface. Our precise atomic-column resolved STEM-EDS composition profiles of the Sc-K, Sn-L, Ba-L and La-L edge signals reveal that the atomic interdiffusion across the interface for both A site (La and Ba) and B site (Sc and Sn) is less than 1 nm (Figure S9). This is similar to that observed in typical STO-based 2DEG studies.[34] Our in-line holography results (Figure 5b) discussed above indicate that most of the carriers do not reside at the cation intermixed region of the BSO, but deeper into the BSO layer. This suggests that intermixing does not dominantly control LSO/BSO 2DEG mobility degradation.

Our LSO/BSO interfacial 2DEG mobility is also below that of La-doped BSO films with similar measured dislocation densities (Figure 4) and similar mobile carrier concentration (Figure 3d).[22] This indicates that dislocation density alone does not completely determine our 2DEG mobilities. For instance, dislocation cores are known to have abundant dangling bonds which effectively scatter electrons in semiconductor films.[35] In fact, comparative STEM studies for both La-doped BSO and undoped BSO films revealed distinctly different local atomic arrangements around dislocation cores.[36] La ions in La-doped BSO can accumulate inside dislocation cores, forming anti-site defects ($La_{Sn}^{-1}$). This will screen the potential attributed to positive charges of the core oxygen vacancies known to reside at dislocation core in perovskite oxides.[37-39], reducing the Coulomb scattering of conduction electrons. This indicates that the defect scattering by dislocation cores may be less in La-doped BSO than in our undoped BSO.[40,41] We conclude that direct comparison of the dislocation density between La-doped BSO films and LSO/BSO interfacial 2DEGs likely does not completely determine the observed differences in electron mobilities, although we have clearly shown that our improved growth technique reduces dislocation density and increases mobility in BSO interfacial 2DEGs. Lattice-matched single crystal substrates[42,43,44] should reduce the dislocation density even further, and are a promising path to achieving the highest mobility in

BSO 2DEGs.

We demonstrated highly mobile 2DEG formation at the LSO/BSO interface with room-temperature mobilities as high as 60 $cm^2/V \cdot s$. Future work will reveal whether the 2DEGs of the LSO/BSO interface can show the exotic interface physics, such as superconductivity,[2,3] two-dimensional hole gas (2DHGs),[29] or quantum hall effect,[45] as have been previously shown in the LAO/STO systems. We anticipate that BSO-based 2DEGs with even higher room-temperature mobilities will be beneficial for transparent field-effect transistor applications as well as a fundamental investigation of new physical phenomena. To this end lattice-matched substrates for BSO interfacial 2DEG heterostructures will provide even more opportunities.

**Experimental Section**

*PLD growth for BSO pseudo-substrate*: 550 nm thick BSO templates were grown on STO (001) substrates by pulsed-laser deposition. Before deposition, STO substrates were treated by a buffered hydrofluoric acid etch and annealed in oxygen at 1000 ℃ for 6 hours to create atomically smooth surfaces with single unit cell steps. The substrate was attached to a resistive heater and positioned ~ 60 mm from the target. A KrF excimer laser (248 nm) was focused on a stochiometric BSO target to an energy density of 1.2 J/cm$^2$ and pulsed at 5 Hz. BSO templates were grown at substrate temperatures of 750 ℃ with an oxygen pressure of 120 mbar, and were slowly cooled down to room temperature under an oxygen pressure of 1 atm. After growth, the BSO film was leached by water for 15 sec to create a $SnO_2$-termination and annealed in 1 atm of oxygen at 1150 ℃ for 2 hours.

*MBE growth of BSO and LSO film*: BSO and LSO thin films were grown in a Veeco GEN10 MBE system. Separate effusion cells containing barium (99.99% purity, Sigma-Aldrich), $SnO_2$ (99.996% purity, Alfa Aesar), lanthanum (99.996% purity, Ames Lab), and scandium (99.9% purity, Alfa Aesar) were heated. The fluxes of the resulting molecular-beams eminating from the effusion cells were measured by a quartz crystal microbalance (QCM) before growth. A commercial ozone generator was used to produce the oxidant molecular beam source (~10 % ozone + 90 % oxygen). The BSO film was grown in an adsorption-controlled regime by supplying an excess $SnO_x$-flux.[22] The background pressure of the oxidant, 10% $O_3$ + 90% $O_2$, was held at a constant ion gauge pressure of $1.0 \times 10^{-6}$ Torr. Subsequently, LSO was grown on top of the BSO film using a layer-by-layer growth method. The fluxes of the La and Sc molecular beams were roughly calibrated by a QCM and then more precisely calibrated by growing $La_2O_3$ and $Sc_2O_3$ binary oxide films and measuring the growth rate using both X-ray reflectivity and *in-situ* reflection high-energy electron diffraction (RHEED) oscillations[46]. For the growth of both the BSO and LSO layers the substrate temperature was maintained between 830 °C and 850 °C, as measured by an optical pyrometer operating at a wavelength of 1550 nm.

*Electrical transport measurement:* Transport measurements used four indium contacts in a van der Pauw geometry in a Quantum Design PPMS between 100 K and 400 K. All resistance measurements were performed by sourcing an alternating dc current and measuring voltages at positive and negative current values. Hall measurements were conducted by sweeping a magnetic field over a range from −20 to 20 kOe. The equations $n_{2D} = I/[(dV_H/dB)q]$ and $\mu = 1/(n_{2D}qr)$ were used to calculate 2D carrier density and mobility, where I is the dc current sourced, $V_H$ is the Hall voltage, q is the electron charge, and r is the sheet resistance. The 3D carrier density was computed using the 2D carrier density and the thickness at which the electron density is zero as determined from the electron holopgraphy profile.

*STEM and EDS measurement:* The cross-sectional sample for (S)TEM measurements was prepared via $Ga^+$ ion beam milling at an accelerating voltage from 30 kV down to 5 kV using a dual-beam focused ion beam system (FIB, Helios 450F1, Thermo Fisher Scientific). An aberration-corrected TEM (JEM-ARM300CF, JEOL) mounting with an energy dispersive x-ray detector was used for TEM based measurements operated at 300 keV. The convergence semi-angle of 23 mrad, and the collection angle ranges of 68–280 mrad, were set for HAADF image and EDS data. During acquisition of the EDS signal, the specimen drift was corrected during observation. Each elemental map is constructed by integrating the signal from Ba-L, Sn-L, La-L, and Sc-K characteristic x-rays, respectively.

*In-line holography measurement:* For inline electron holography, a through-focal series of TEM bright field images was acquired using a 2 k × 2 k CCD camera (UltraScanXP 100FT, Gatan Inc.), varying the defocus values from -1800 nm to 1800 nm in 600 nm steps. An energy filter (Quantum spectrometer ER965, Gatan Inc.) was used to remove inelastically scattered electrons outside an energy window of 0 ± 5 eV. An objective aperture with a diameter of 20 μm was used to select the transmitted beam for bright field imaging. To minimize the effects of dynamical diffraction, all images were taken at diffraction condition off the $[1\bar{1}0]$ zone axis with the interface being kept in an end-on projection.[47] The obtained inline electron holograms were used to reconstruct the phase shift of the transmitted beam using the full resolution wave reconstruction (FRWR) algorithm.[48] The reconstructed phase images were converted into the map of the projected electrostatic potential by assuming the phase-object approximation for a non-magnetic material. The charge-density map was obtained from the potential data using Poisson's equation.

**Supporting Information**

Supporting Information is available from the author.

**Acknowledgements**

This research is funded in part by the Gordon and Betty Moore Foundation's EPiQS Initiative, grant GBMF9065 to C.-B.E., Vannevar Bush Faculty Fellowship (ONR N00014-20-1-2844) AFOSR (FA9550-15-1-0334). Transport measurement at the University of Wisconsin–Madison was supported by the US Department of Energy (DOE), Office of Science, Office of Basic Energy Sciences (BES), under award number DE-FG02-06ER46327. This work was supported in part by the National Science Foundation (Platform for the Accelerated Realization, Analysis, and Discovery of Interface Materials (PARADIM)) under Cooperative Agreement No. DMR-2039380. The work at SKKU was supported by the Samsung Research Funding and Incubation Center of Samsung Electronics under Project Number SRFC-MA1702-01.

**Conflict of Interest**

The authors declare no conflict of interest.

**Data availability Statement**

The data that support the findings of this study are available from the corresponding author on reasonable request.

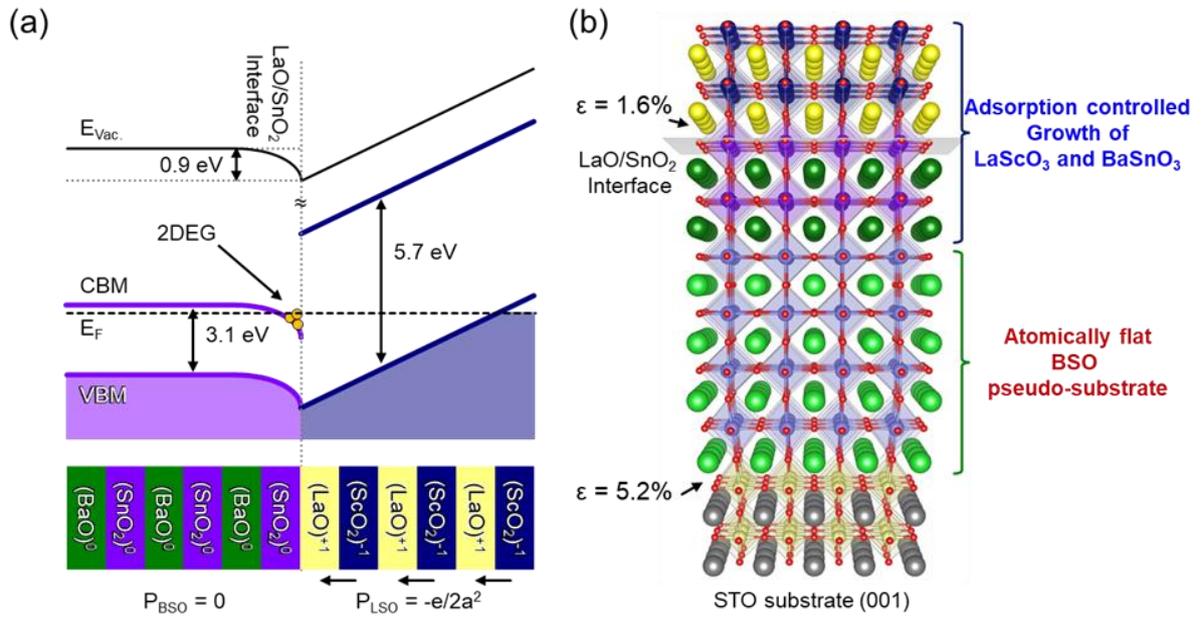

**Figure 1.** Experimental design for 2DEG formation at the LSO/BSO interface. a) The band diagram of LSO/BSO heterostructures above the critical thickness, showing 2DEG formation at the interface. b) Schematic of a BSO pseudo-substrate (550 nm thick BSO film on STO (001) substrate) and our strategy to acquire high quality 2DEGs at the LSO/BSO interface. This process minimizes the dislocation scattering centers in the MBE-grown BSO 2DEG channel layer on an atomically flat $SnO_2$-terminated annealed BSO pseudo-substrate. The annealing treatment reduces the dislocation density and produces $SnO_2$-terminated atomically flat surfaces. The structure consists of MBE-grown LSO (several unit cells) on top of the adsorption-controlled BSO thin film (45 nm) that is also grown by MBE on the BSO pseudo-substrate.

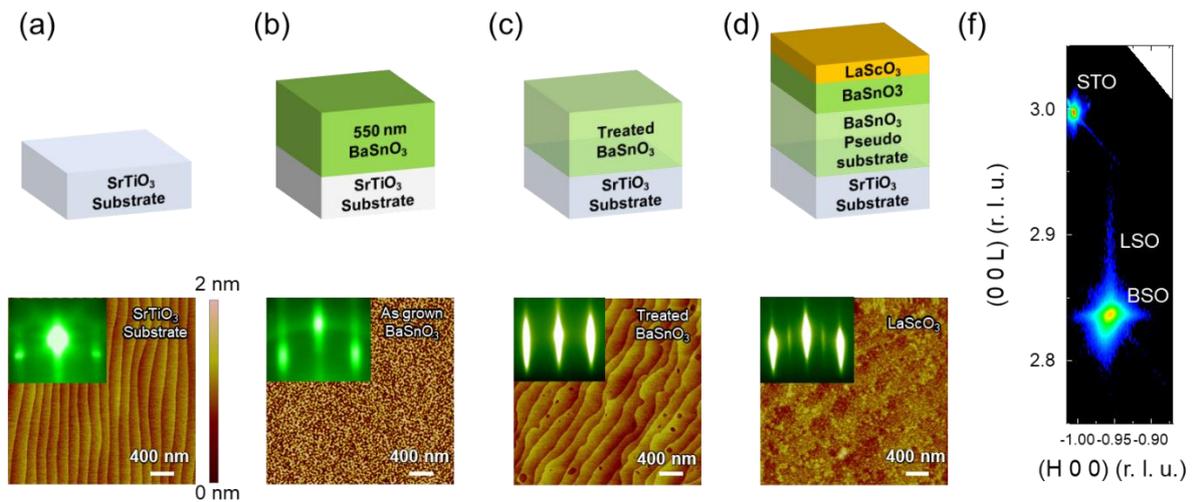

**Figure 2.** Fabrication of a high mobility 2DEG at the LSO/BSO interface, showing AFM and RHEED (bottom) after each step of the fabrication (top). a) STO (001) substrate, b) as-grown 550 nm thick BSO film (BSO pseudo-substrate), c) after water leaching and thermal annealing of BSO pseudo-substrate, and d) LSO (10 u.c.) / BSO (45 nm) grown on the BSO pseudo-substrate via MBE. The insets in the AFM images represent the RHEED patterns at each step of the PLD (steps a) and b)) and MBE growths (steps c) and d)). f) Reciprocal space mapping (RSM) around 103 reflections from LSO (10 u.c.) / BSO (45 nm) grown on the BSO pseudo-substrate.

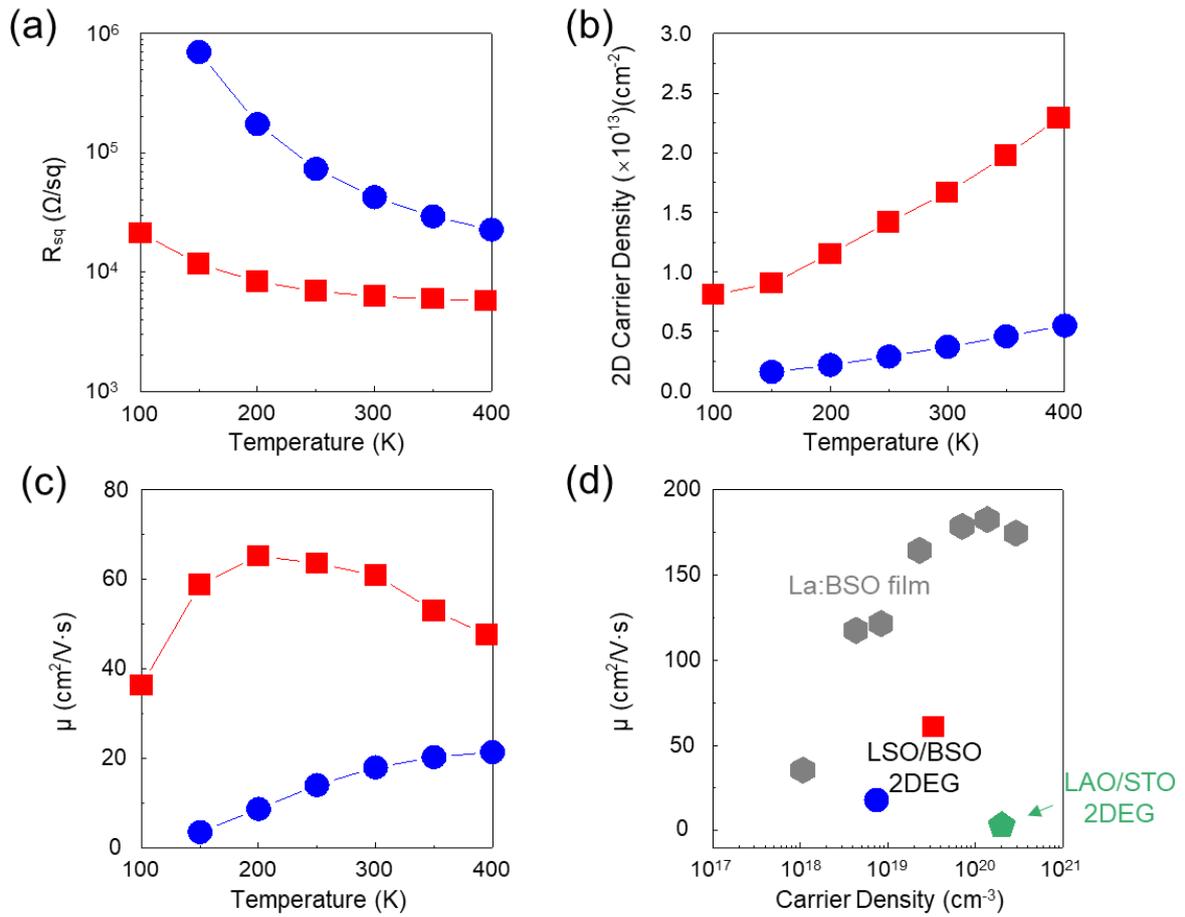

**Figure 3.** Transport properties of the LSO/BSO heterostructures. a) Sheet resistance, b) carrier density and c) mobility of LSO (10 u.c.) / BSO (60 nm) directly grown on an STO (001) substrate (closed blue circle) and LSO (10 u.c.) / BSO (45 nm) grown on the BSO pseudo-substrate (closed red square). d) Electron mobility at 300 K as a function of carrier density for 2DEGs at oxide heterointerfaces in this work, La:BSO film[22] and LAO/STO[29] reported in the literature.

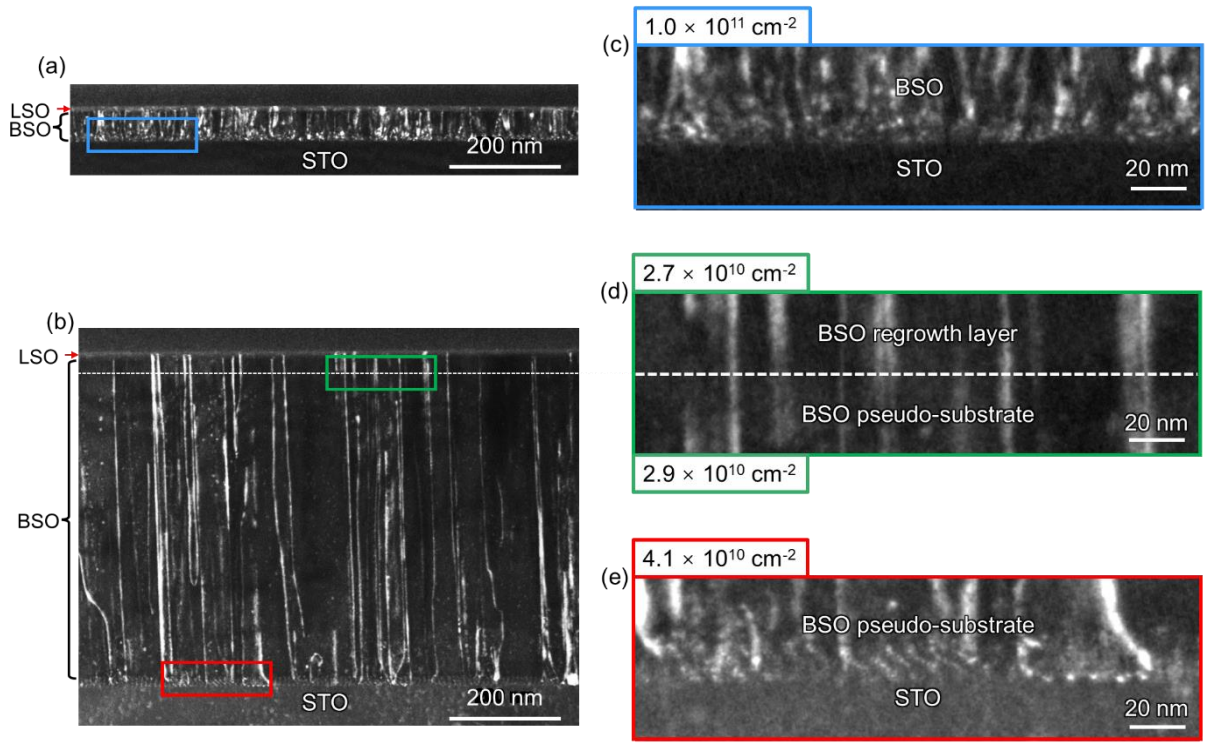

**Figure 4.** Threading dislocation density approximated from the cross-sectional weak beam dark field images. Weak beam dark field images of a) LSO (10 u.c.) / BSO (60 nm) / STO without a BSO pseudo-substrate and b) LSO (10 u.c.) / BSO (45 nm) grown on the BSO pseudo-substrate. c) Magnified image at the BSO/STO interfaces of LSO (10 u.c.) / BSO (60 nm) / STO. d) Interface between the BSO pseudo-substrate and BSO regrowth layer, indicating little change in defect density. e) Magnified image at LSO (10 u.c.) / BSO (45 nm) grown on the BSO pseudo-substrate and extracted dislocation densities showing reduced dislocation density as a result of high temperature annealing .

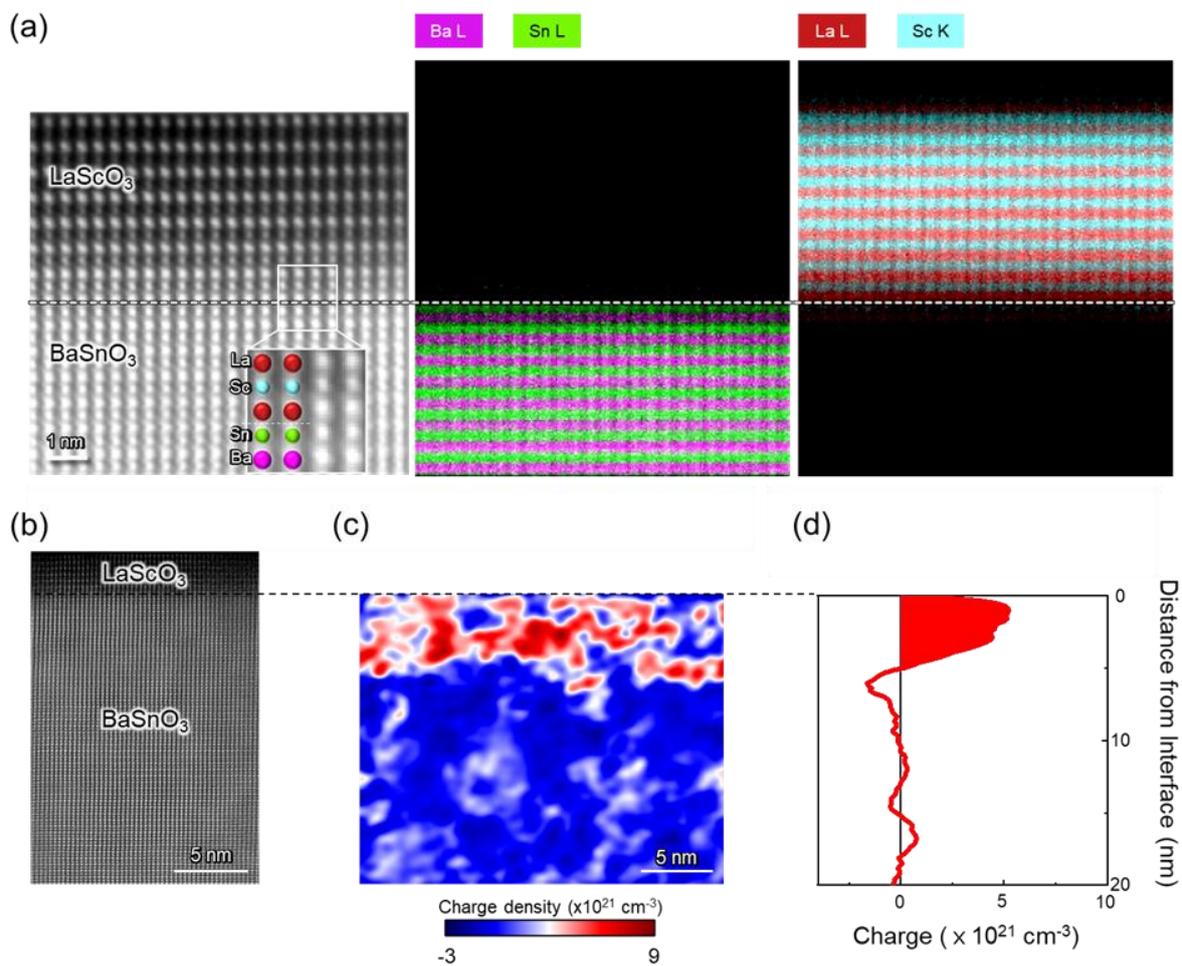

**Figure 5.** Elemental analysis and electron distribution at the LSO/BSO interface. a) Atomic resolution STEM-HAADF images and EDS elemental mapping of the LSO/BSO interface, indicating an atomically abrupt interface. b) Direct imaging of the LSO/BSO interface by STEM HAADF imaging are shown next to c) the charge density maps and d) 1D electron density profiles obtained by in-line electron holography for the LSO (10 u.c.) / BSO (90 nm) grown on an STO substrate.

Supporting Information for

**Oxide two-dimensional electron gas with high mobility at room-temperature**

Kitae Eom,[1] Hanjong Paik,[2,3] Jinsol Seo,[4] Neil Campbell,[5] Evgeny Y. Tsymbal,[6] Sang Ho Oh,[4] Mark Rzchowski,[5] Darrell G. Schlom,[2,7,8] and Chang-Beom Eom*,[1]

**Contents**

**Section S1. The effect of post treatment on the BSO pseudo-substrate**
Figure S1. Structural characterization of PLD-grown BSO thin films before and after thermal treatment.

**Section S2. MBE growth of BSO films**
Figure S2. RHEED intensity oscillations during the MBE growth of BSO films

**Section S3. Dislocation density analysis**
Figure S3. A magnified STEM-HAADF image near the LSO/BSO interface regions along the [110] zone axis.
Figure S4. Dislocation density calculations using weak beam dark field images
Figure S5. STEM-HAADF images and geometric phase analysis of the BSO/STO interface

**Section S4. Critical thickness threshold for conductivity at the LSO/BSO interface**
Figure S6. Thickness-dependent evolution of *in-situ* RHEED intensity oscillations during the PLD growth of LSO films.
Figure S7. Room temperature sheet conductivity as a function of the number of LSO unit cells.

**Section S5. Inline holography measurement**
Figure S8. Charge distribution in the LSO (4 u.c.)/BSO interface
Figure S9. Atomic-column resolved STEM-EDS composition profiles

*Correspondence should be sent to eom@engr.wisc.edu.

**Section S1. Structural characterization of PLD-grown BSO thin films before and after thermal treatment. The effect of post treatment on the BSO pseudo-substrate.**

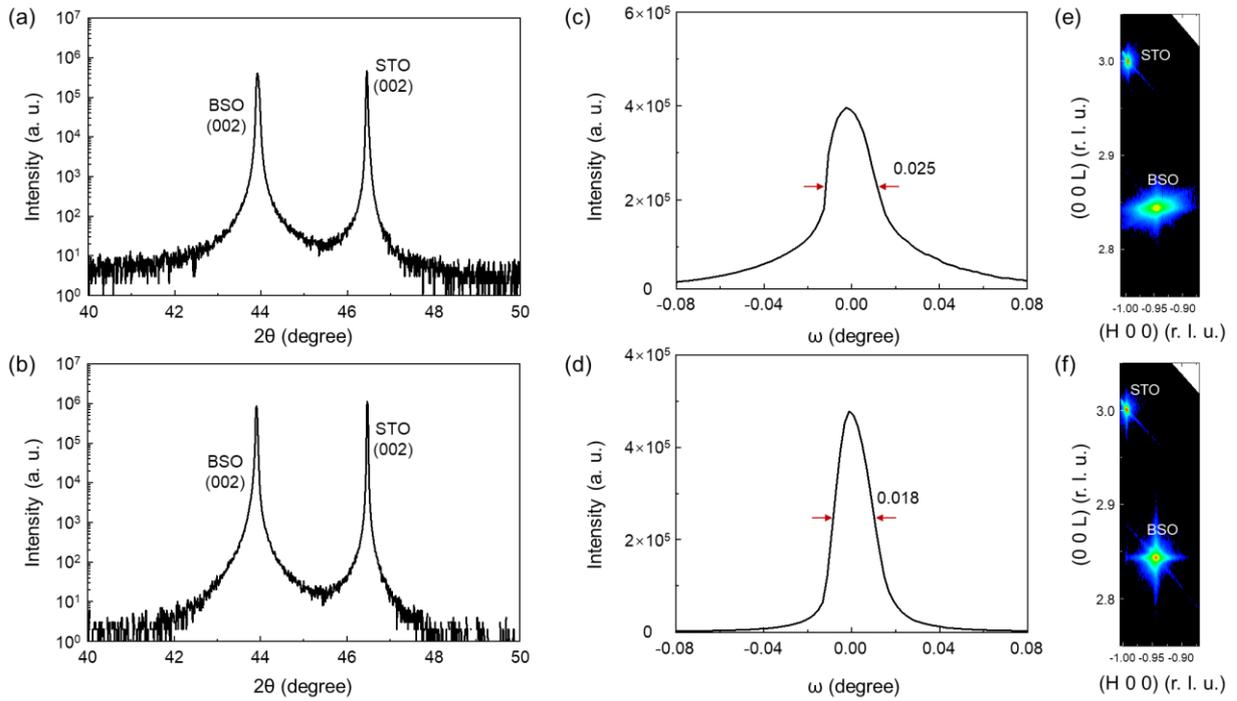

**Figure S1.** Structural characterization of PLD-grown BSO thin films before and after thermal treatment. $\theta$-$2\theta$ scans (a) before and (b) after post treatment. Rocking curves of the BSO 002 peak (c) before and (d) after post treatment. The changes of the full width at half max (FWHM) of the BSO 002 and corresponding RSM result taken around the STO 103 reflection (e) before and (f) after post treatment reflects the dislocation annihilation effect.

**Section S2. MBE growth of BSO films**

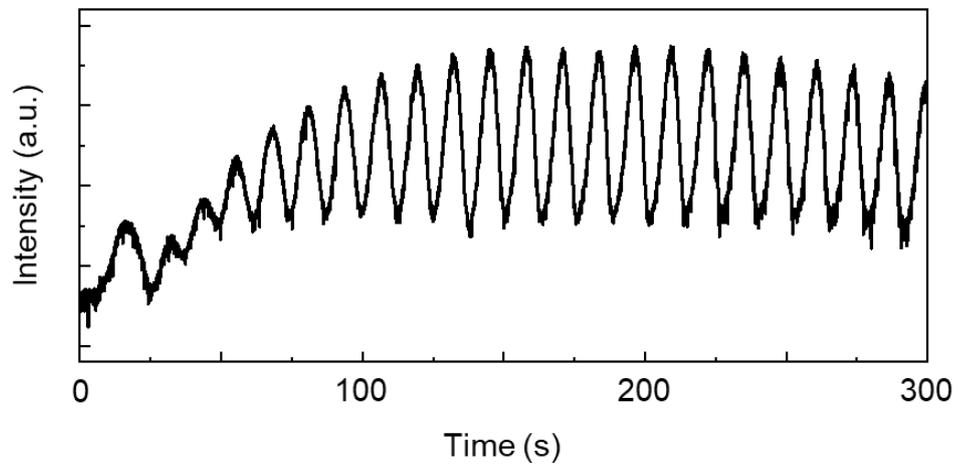

**Figure S2.** RHEED intensity oscillations during the MBE growth of BSO films.

**Section S3. Dislocation density analysis**

The threading dislocation density was measured by counting dislocation lines in TEM weak beam dark field images following the equation below,

$$\rho = \frac{N}{Lt} \quad (0.1)$$

where *N* is the total number of dislocations in a field of view, *L* is the total length of the field of view and *t* is the thickness of the TEM specimen. The weak beam dark field images were taken by selecting the in-plane 002 beam. The total field of view, *L* of LSO (10 u.c.)/BSO (45 nm)/BSO (550 nm) and LSO (10 u.c.)/BSO (60 nm) specimens were 3.629 $\mu$m and 3.057 $\mu$m, respectively. To measure the total number of dislocations, *N*, a few lines were drawn in the in-plane direction, and the number of intersections with the dislocations was counted and averaged. Considering that the TEM specimen is wedge-shaped and that threading dislocations can escape through the surface of a thin TEM specimen, the number of representative dislocations was counted near the bottom interfaces (BSO/STO interface). The number of dislocations counted are 128 and 332 for LSO (10 u.c.)/BSO (45 nm)/BSO (550 nm) and LSO (10 u.c.)/BSO (90 nm) specimens, respectively. The thickness of specimens, *t* determined by using the EELS log-ratio method,[1] were 110.67 nm and 98.87 nm on average with small thickness gradients for the LSO (10 u.c.)/BSO (45 nm)/BSO (550 nm) and LSO (10 u.c.)/BSO (60 nm) specimens, respectively.

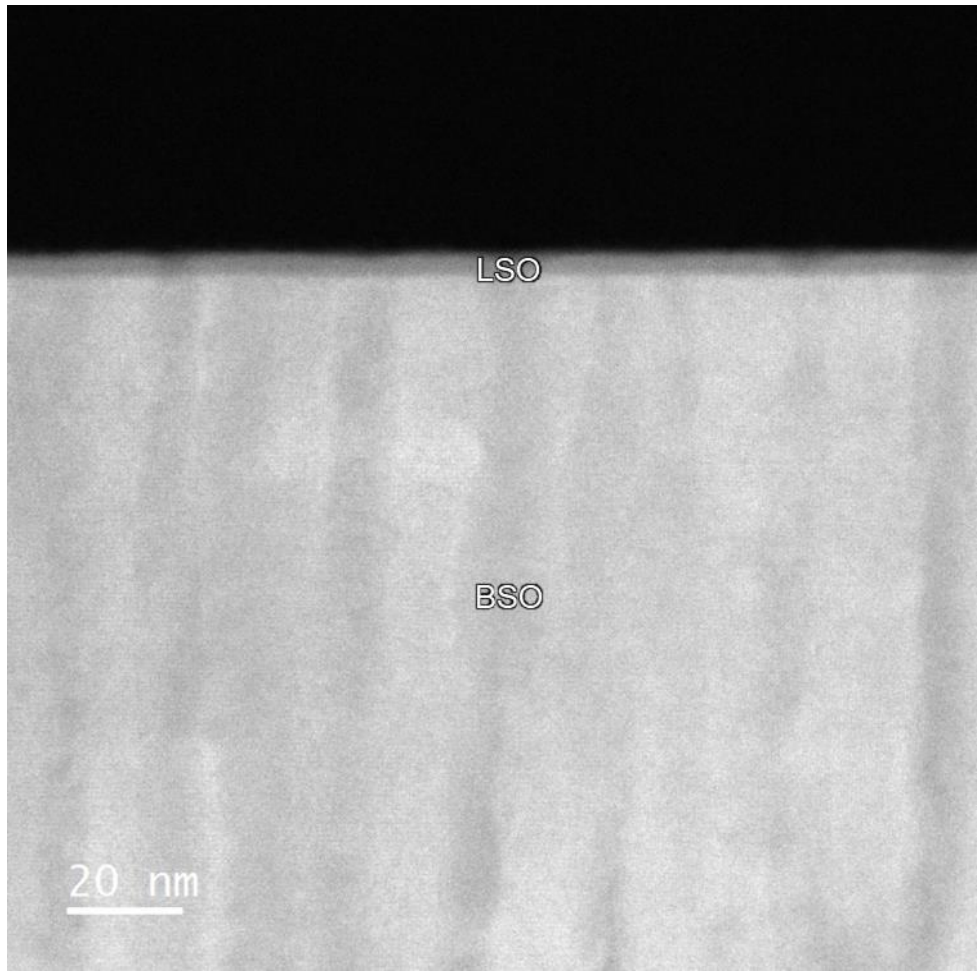

**Figure S3.** A magnified STEM-HAADF image near the LSO/BSO interface regions with a zone axis of [110]. Threading dislocations propagate in the film growth direction from the BSO/STO interface into the LSO layer.

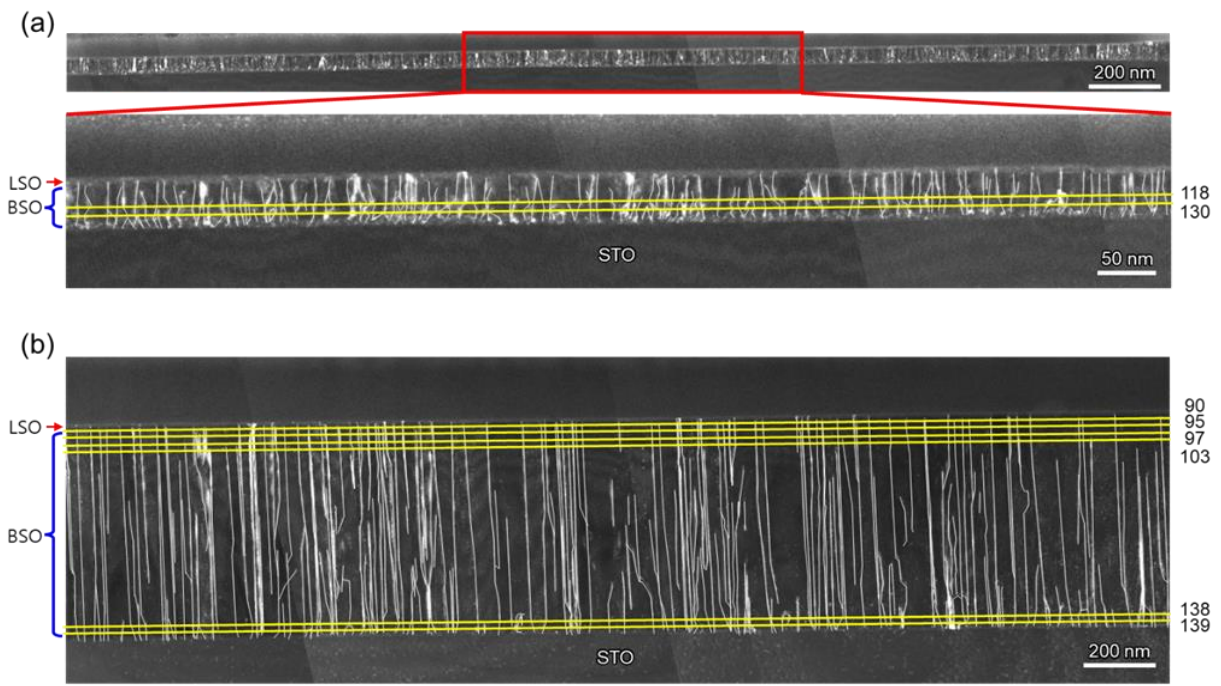

**Figure S4.** Dislocation density calculations made by using weak beam dark field images from (a) LSO (10 u.c.) / BSO (60 nm) / STO without the BSO pseudo-substrate and (b) LSO (10 u.c.) / BSO (45 nm) grown on the BSO pseudo-substrate. An in-plane 200 beam was excited for imaging. Several images are stitched together for both samples due to the limited field of view. Dislocations are highlighted with white lines. To extract dislocation densities, we drew lines parallel to the film (yellow lines). The intersections were counted, and averaged numbers were used to estimate the dislocation density. The total fields of view are 3.1 μm and 3.6 μm for LSO (10 u.c.) / BSO (60 nm) / STO and LSO (10 u.c.) / BSO (45 nm) grown on a BSO psuedo-substrate, respectively.

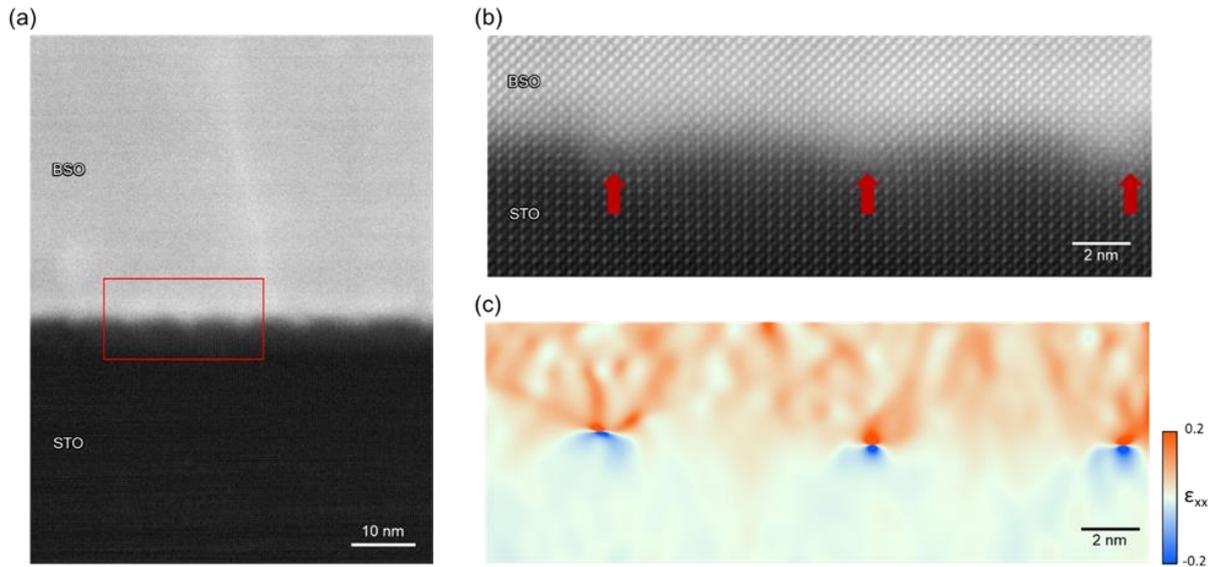

**Figure S5.** STEM-HAADF images and geometric phase analysis on the BSO/STO interface at the LSO (10 u.c.)/BSO (45 nm) grown on a BSO pseudo-substrate. (a) STEM HAADF image with low magnification. (b) atomic-column resolved STEM-HAADF image from the area marked with the red rectangle in (a). The red arrows indicate misfit dislocations. (c) In-plane strain ($\varepsilon_{xx}$) map from GPA analysis, which is obtained in the same area as (b).

**Section S4. Critical thickness threshold for conductivity at the LSO/BSO interface**

To investigate the critical thickness threshold for conductivity at the LSO/BSO interface, LSO films were grown by PLD on MBE-grown 90 nm thick BSO/STO films. The LSO target was ablated using a KrF (248 nm) excimer laser at a repetition rate of 3 Hz and with a fluence of 1.8 J/cm$^2$. The substrate to target distance was 62 mm. The LSO films were grown at a temperature of 750 ℃ with an oxygen pressure of 10 mbar, and were slowly cooled down to room temperature under an oxygen pressure of 1 atm. Before the growth, the BSO/STO film was leached by DI water for 15 sec to obtain an SnO$_2$-terminated surface.[2] The LSO film thickness is controlled by monitoring the RHEED oscillations (Figure S6).

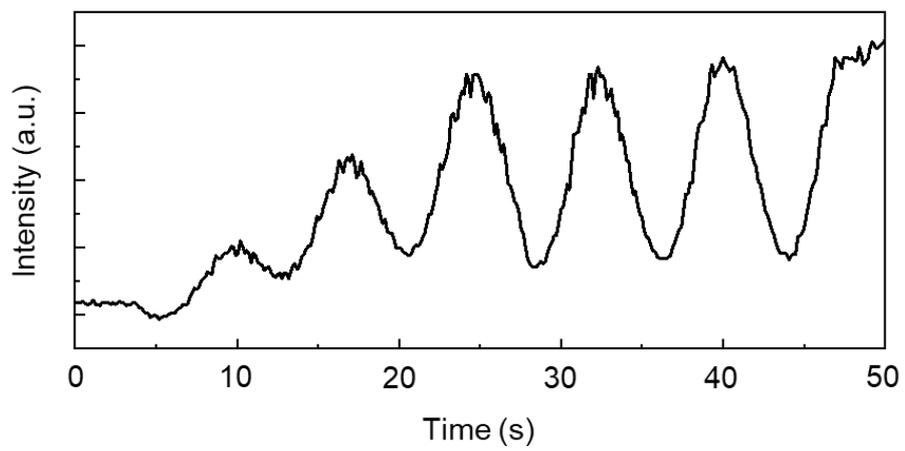

**Figure S6.** Thickness-dependent evolution of *in-situ* RHEED intensity oscillations during the PLD growth of LSO films.

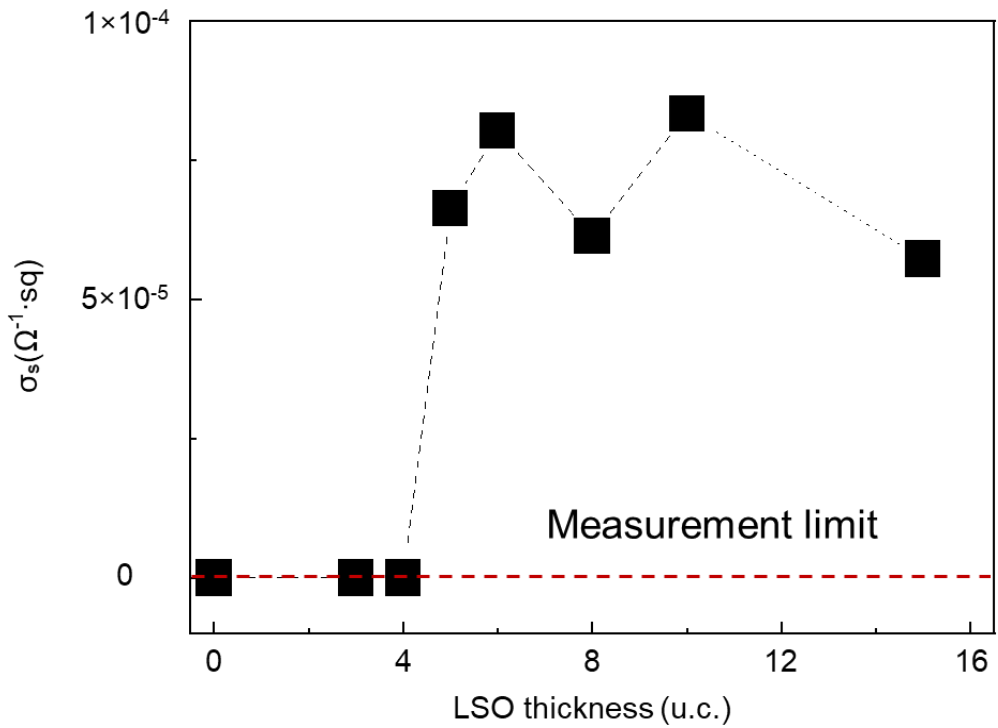

**Figure S7.** Sheet conductivity at room temperature as a function of the number of LSO unit cells. LSO layers were grown by PLD on 90 nm thick BSO / STO (001) samples that had been grown by MBE and leached by water to make $SnO_2$-terminated surfaces. This plot indicates that a conducting interface is formed when the thickness of the LSO layer is more than 4 u.c., clearly showing the LSO critical thickness limit for forming a conducting 2DEG interface on BSO.

**Section 5. Inline holography measurement**

In our previous report, it has been demonstrated that the inline holography technique is capable of measuring highly confined electron 2DEGs even at interfaces having a low spatial frequency issue.[3] The charge density map/profile should be carefully considered because of the mean inner potential and a limited spatial resolution of the phase image. The estimated mean inner potentials of LSO and BSO are 17.1 and 16.3, respectively[4], which may cause a small artificial peak at the interface in the charge density maps and profiles (Figure 5b and Figure S8). Nevertheless, it is still reasonable to conclude distinctly different charge confinement at the LSO (10 u.c.) / BSO (90 nm) interface compared to that of the LSO (4 u.c.) / BSO (90 nm) interface.

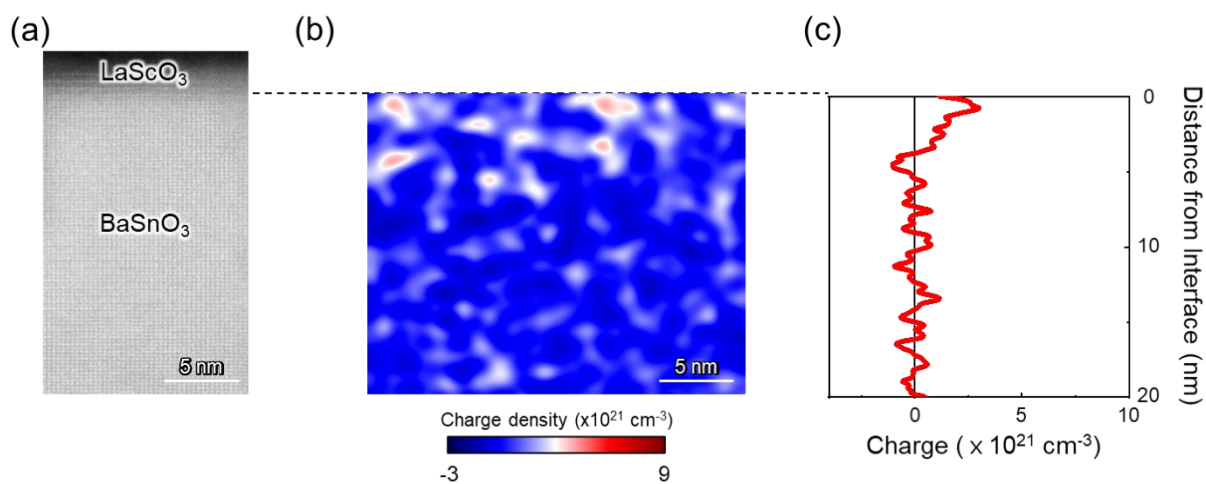

**Figure S8.** Charge distribution in the LSO (4 u.c.)/BSO interface. (a) STEM HAADF images are shown next to (b) the charge density maps and (c) 1D electron density profiles obtained by in-line electron holography for the LSO (4 u.c.) / BSO (90 nm) grown directly on an STO substrate.

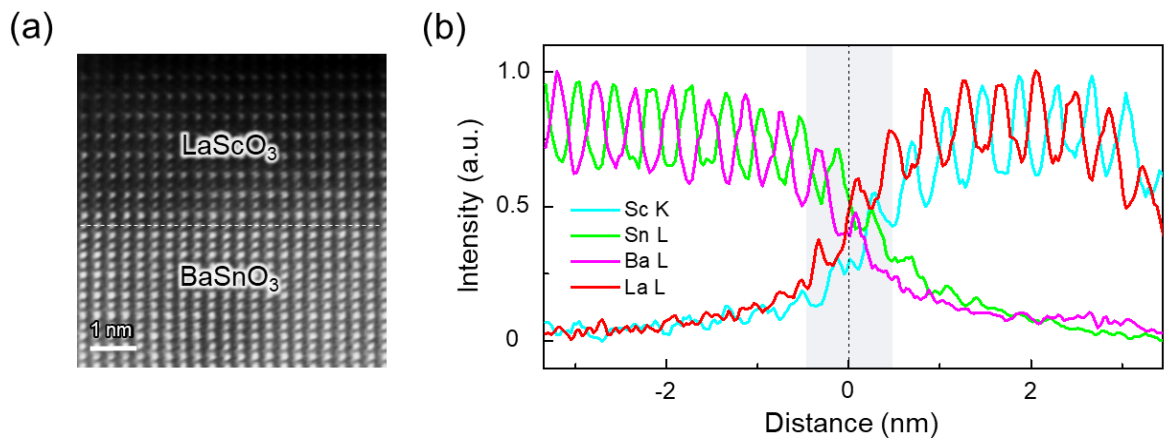

**Figure S9.** Atomic-column resolved STEM-EDS composition profiles. (a) Atomic resolution HAADF STEM image of the LSO (10 u.c.) / BSO (90 nm) / STO along the [001] zone axis of STO (b) normalized intensity profiles of STEM-EDS obtained at the LSO/BSO interface, which is constructed by selecting Sc-K, Sn-L, Ba-L and La-L characteristic X-rays, respectively.